\documentclass[aps,prb,twocolumn,superscriptaddress,floats]{revtex4}
\usepackage{txfonts}
\usepackage{amssymb}
\usepackage{graphicx}
\usepackage{sidecap}

\begin{document}

\title{Critical role of substrate in the high temperature  superconductivity of  single layer FeSe on Nb:BaTiO$_3$}

\author{R. Peng}\thanks{These authors contributed equally to this work.}
\author{H. C. Xu}\thanks{These authors contributed equally to this work.}
\author{S. Y. Tan}
\author{M. Xia}
\author{X. P. Shen}
\author{Z. C. Huang}
\author{C. H. P. Wen}
\author{Q. Song}
\author{T. Zhang}
\author{B. P. Xie}
\author{D. L. Feng}\email{dlfeng@fudan.edu.cn}

\affiliation{State Key Laboratory of Surface Physics,  Department of Physics, and Advanced Materials Laboratory, Fudan University, Shanghai 200433, People's Republic of China}

\date{\today}

\maketitle

\textbf{In the quest for high temperature superconductors, the  interface between a metal and a dielectric   was proposed  to possibly achieve very high superconducting transition temperature ($T_c$) through interface-assisted pairing \cite{Little,Ginzburg}. Recently, in single layer FeSe (SLF) films grown on SrTiO$_3$  substrates, signs for  $T_c$ up to  65~K  have been reported \cite{FeSeZhou2,TanFeSe}.  However, besides doping electrons and imposing strain, whether and how the substrate facilitates the superconductivity are still unclear. Here we report the growth of various SLF films  on thick BaTiO$_3$ films atop KTaO$_3$ substrates, with  signs for  $T_c$ up to $75$~K, close to the liquid nitrogen boiling temperature.  SLF of similar doping and lattice is found to  exhibit high $T_c$ only if it is on the substrate, and its band structure  strongly depends on the substrate. Our results highlight the profound role of substrate on the high-$T_c$ in SLF, and provide new clues for understanding its mechanism.
}

The record $T_c$ has long been 56~K for the bulk iron-based high temperature superconductors (Fe-HTS's).
Recently,  in SLF films grown on   Nb doped SrTiO$_3$ (NSTO) substrate (hereafter referred as FeSe$^S$) by molecular beam epitaxy (MBE),  both scanning tunneling spectroscopy and angle-resolved photoemission spectroscopy (ARPES) experiments have observed large superconducting gap \cite{FeSeXue,FeSeZhou}, which  closes  at $\sim$65~K (ref.~\onlinecite{FeSeZhou2,TanFeSe}).
On the other hand, \textit{ex situ} transport measurements of SLF films capped with FeTe and silicon show an onset $T_c$ of about 40~K, with  signs for a possible Berezinskii-Kosterlitz-Thouless (BKT) transition \cite{WangjianFeSe}. Although whether this is the true $T_c$ remains to be checked by \textit{in-situ} transport measurements, it is likely that the phase fluctuation in this two dimensional system is  strong, so that the ARPES  gap-closing temperature, $T_g$,  only measures the Cooper pair formation temperature,  instead of $T_c$. Nevertheless,
it implies  that a  $T_c$ equivalent to  $T_g$   can be achieved if the fluctuations in such a two dimensional system could be  suppressed.

Based on the existing experiments, the NSTO substrate affects SLF in the following ways. 1. The oxygen vacancies in the substrate induces extra electrons that are  transferred to the interfacial FeSe layer \cite{TanFeSe,ZhangSBcalc}, which suppresses the antiferromagnetism in SLF and allows for a high $T_g$ (ref.~\onlinecite{TanFeSe}).
2. The in-plane lattice constant ($a$) of NSTO is  3.905~\AA~, thus it imposes a large tensile strain on SLF, noting $a$=3.765~\AA~ for bulk FeSe. It was shown that the tensile strain would enhance the  antiferromagnetic interactions  in FeSe (ref.~\onlinecite{CaoFeSe}).  However,  for SLF films with extremely expanded $a$ of $\sim$3.99~\AA~ in epitaxially grown FeSe/NSTO/KTaO$_3$ heterostructures (hereafter referred as  FeSe$^{SX}$),  the $T_g$ increases just moderately to about 70~K (ref.~\onlinecite{PengSTOKTO}).
3.  It was reported recently that electrons in FeSe might interact with an $\sim 80$~meV optical phonon in the substrate \cite{ZXShenFeSe}.  So far, it is still unclear whether and how the substrate plays a more direct role in mediating the superconducting pairing in SLF.

\begin{figure*}[t]
 \includegraphics[width=17cm]{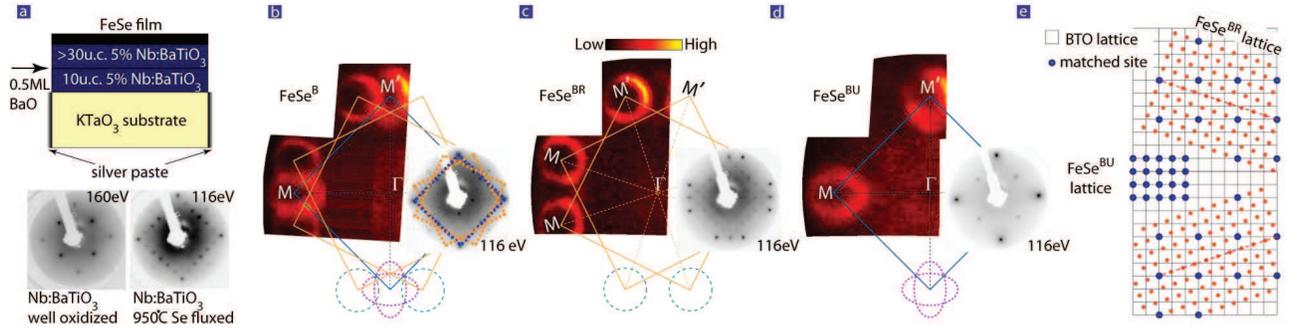}
\caption{\textbf{The structure and Fermi surface of FeSe/Nb:BTO/KTO  film.} \textbf{a}, The schematic cross-section of the FeSe/Nb:BTO/KTO  film. The LEED (low energy electron diffraction) patterns for the Nb:BTO/KTO film before and after the heat treatment under Se flux are shown on the bottom. To prepare for FeSe$^{BU}$ growth,  an extra 0.5~ML of BaO is inserted to eliminate the 3$\times$3 superstructure as indicated.
\textbf{b},  Left: the Fermi surface of a  film named FeSe$^{B}$ as represented by the photoemission intensity map at the Fermi energy. The intensity was integrated over a window of ($E_F$, $E_F$-20~meV).  Right:  its LEED pattern. Besides the main spots that reflect the unrelaxed lattice of the substrate (inicated by the blue square), there are spots corresponding to two $\pm 18.5^{\circ}$ rotated lattices  (indicated by the yellow squares).
Correspondingly, there are three sets of Fermi surfaces, whose respective Brillouin zone  are derived based on the high symmetry points.
\textbf{c},  Left: the Fermi surface of a film named FeSe$^{BR}$ resulting from post-annealing FeSe$^{B}$, where only two sets of Fermi surfaces are observed,  corresponding to the two $\pm 18.5^{\circ}$ rotated lattices. Right:   its LEED pattern, where the main spots are weakened, and the spots for the rotated lattice are enhanced.
\textbf{d}, Left: The Fermi surface of a film named FeSe$^{BU}$, where only the elliptical Fermi surface sheets are observed.
Right:   its LEED pattern. No superstructure spots are observed, and the lattice is unrelaxed.
\textbf{e}, Sketches for the lattices of various FeSe/Nb:BTO/KTO  films.}
\label{BTObasic}
\end{figure*}

To  study the role of substrate in SLF superconductivity, different substrates should be examined.
BaTiO$_3$ (BTO)  can be made with the same TiO$_2$ termination as STO.  However, due to the spontaneous polarization of BTO, atomically flat surface is hard to obtain with bulk BTO.
To overcome this, since BTO lattice ($a$=3.992~\AA)  is  merely 0.075~\% larger than that of KTaO$_3$ (KTO), we first grew $>$40 u.c. of Nb-doped BTO (NBTO, to provide a conducting substrate) on the  (001)  surface of a KTO crystal by ozone assisted MBE, then grew SLF on the NBTO film (Fig.~\ref{BTObasic}a) as described in the Methods section.
This heterostructure, hereafter referred as  FeSe$^B$, is then transferred to the analysis chamber under ultra high vacuum for the  \textit{in-situ} measurements of low energy electron diffraction (LEED) and  ARPES.

 \begin{figure*}
\includegraphics[width=15cm]{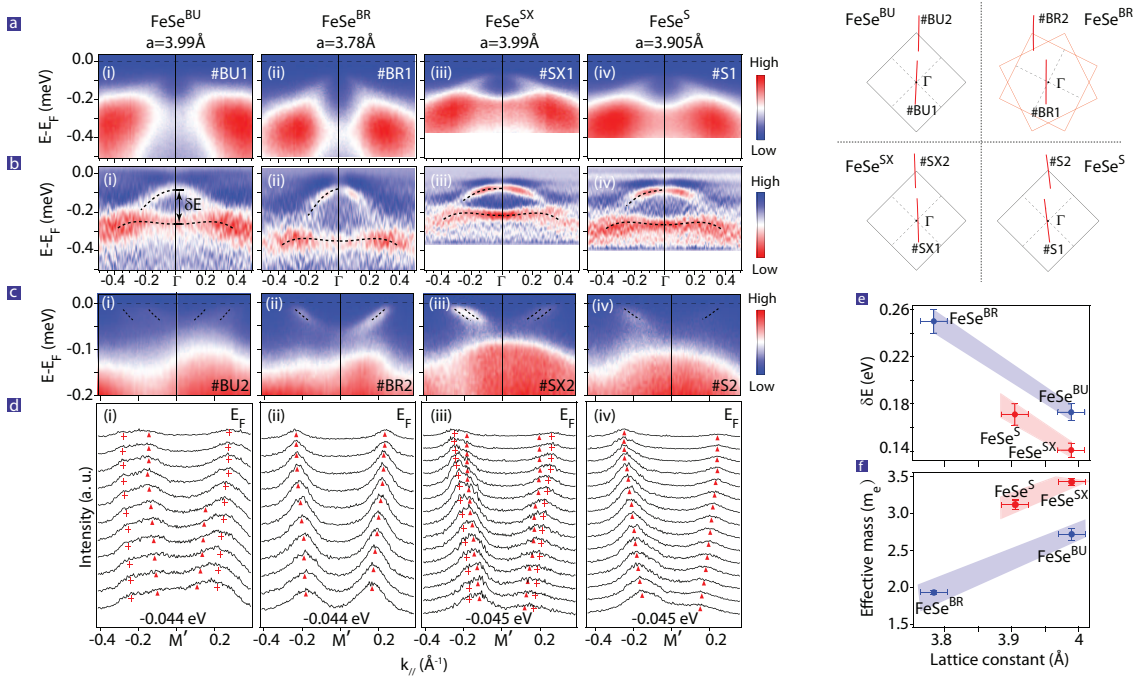}
\caption{\textbf{The band structure of various FeSe films.} \textbf{a,b}, The photoemission intensity (\textbf{a}(i)-(iv))  and  the corresponding second derivative with respect to energy to highlight the dispersions (\textbf{b}(i)-(iv)) across the $\Gamma$ for FeSe$^{BU}$, FeSe$^{BR}$, FeSe$^{SX}$, and FeSe$^{S}$, respectively.
\textbf{c,d}, The photoemission intensity across M (\textbf{c}(i)-(iv)), and the  corresponding momentum distribution curves (MDC's) (\textbf{d}(i)-(iv)) for these films. \textbf{e} The  separation between bands   at $\Gamma$ as defined in  \textbf{b}(i)  as a function of $a$. \textbf{f}, effective mass of  the parabolic band  at $\Gamma$ as a function of $a$. The top-right inset illustrates the momentum locations for data in \textbf{a}-\textbf{d}.  We note that the slight rotation of the cuts  does not affect the quantitative conclusions here (see Supplementary Information).  The $\#$BU1 and $\#$BR1 data were taken at 45~K, while others were taken at 30~K.
}
\label{dispersion}
\end{figure*}

\begin{figure}[t]
 \includegraphics[width=8.5cm]{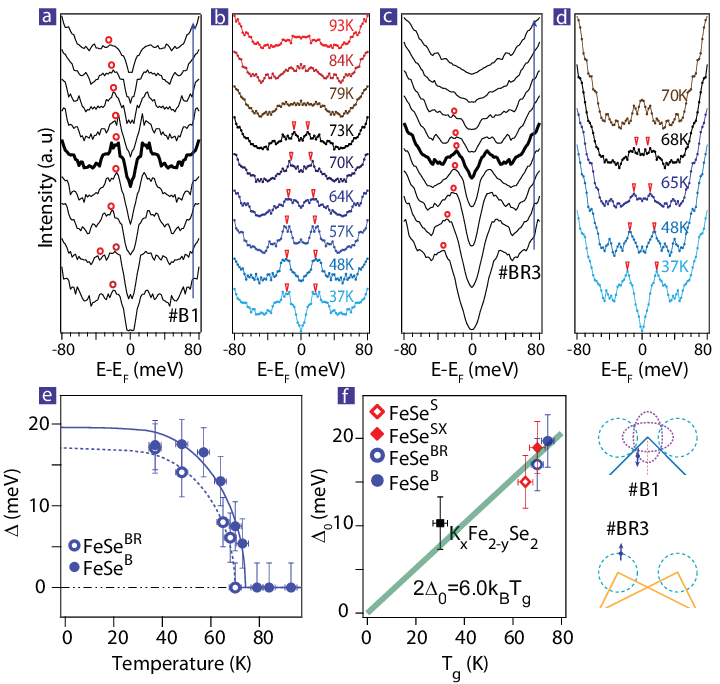}
\caption{  \textbf{Temperature dependence of the superconducting gap for FeSe films.} \textbf{a},  The symmetrized energy distribution curves (EDC's)  for FeSe$^{B}$ along cut~\#B1 as indicated in the inset. The circles mark the dispersion. The thick EDC curve is at the normal state Fermi momentum ($k_F$). \textbf{b}  Temperature dependence of the symmetrized EDC at the marked  $k_F$ for FeSe$^{B}$. The coherent peak positions are marked by triangles. \textbf{c,d}, similar as \textbf{a,b}, but for FeSe$^{BR}$ film.
 \textbf{e},  The  superconducting gap determined as a function of temperature from data in \textbf{b} and \textbf{d}.  The solid lines are the fit for the gap-temperature dependence based on BCS theory. \textbf{f}, The maximal low temperature gap versus $T_g$ plot, including the values for   K$_x$Fe$_{2-y}$Se$_2$  bulk (ref.~\onlinecite{ZhangNM}), FeSe$^{S}$ (ref.~\onlinecite{TanFeSe})  and FeSe$^{SX}$ (ref.~\onlinecite{PengSTOKTO}). Note for  FeSe$^{SX}$ and FeSe$^{B}$, the maximal  values of the anisotropic gap distributions are used (see Supplementary Information).}
\label{Tdep}
\end{figure}

The as-grown well-oxidized NBTO film exhibits a nice surface with the symmetry of the substrate, as shown by its LEED pattern in Fig.~\ref{BTObasic}a. However, after heat treated at 950$^{\circ}$C under the selenium flux, the NBTO film exhibits  a  3$\times$3  reconstruction. This has an direct impact on the SLF grown on it.
The left part of Fig.~\ref{BTObasic}b shows the photoemission intensity map at the Fermi energy ($E_F$) for  FeSe$^B$, where three sets of Fermi surfaces are observed.
 One set (unrotated) is made of elliptical Fermi surface sheets, similar to that observed in FeSe$^{SX}$ (ref.~\onlinecite{PengSTOKTO}). The other two sets are made of more circular sheets, similar to the Fermi surface topology of FeSe$^{S}$ film.   Although these Fermi surface sheets appear to cross each other, their band structures do not exhibit any hybridization. This indicate that they are likely originated from three types of spatially separated domains. Moreover, the superstructure in the substrate does not cause any observable folding, indicating that the crystal potential of the substrate affect the FeSe rather weakly \cite{OuHongweiPRL}.
  The in-plane lattice constant $a$ of each domain could be derived from the inversed Brillouin zone size determined by high symmetry points of photoemission map \cite{TanFeSe}.
   We found that one domain preserves the lattice  of the KTO substrate ($a\sim$3.99~\AA~), and the other two domains  rotated about $\pm 18.5^{\circ}$ with   $a\sim$3.78~\AA~.
Consistently,  the LEED pattern of  FeSe$^B$ in  Fig.~\ref{BTObasic}b can be decomposed into three sets of spots.
Besides the main spots that reflect the unrotated lattice of the substrate (highlighted by the blue square), there are spots corresponding to two $\pm 18.5^{\circ}$ rotated lattices (highlighted by the two yellow squares).  As shown in Fig.~\ref{BTObasic}e, these three domains of FeSe match the reconstructed substrate. Particularly,  every 10 periods of the rotated FeSe lattice match the diagonal of a 3$\times$9 rectangle of the NBTO lattice (the dashed line in Fig.~\ref{BTObasic}e).

After post-annealing for over 15 hours (details are shown in the Supplementary Information), the $\pm 18.5^{\circ}$ domains with the relaxed lattice could be further enhanced (such a film is referred as FeSe$^{BR}$ hereafter).  As shown in Fig.~\ref{BTObasic}c, the spots correspond to the rotated lattices become stronger, while those correspond to the unrelaxed lattice is weakened. As expected, the elliptical Fermi surfaces sheets is neligibly weak.
Intriguingly, we found that if some extra BaO is inserted during the growth of NBTO film, as shown in Fig.~\ref{BTObasic}a,  the resulting FeSe film only contains single unrelaxed domain, as illustrated by its  LEED pattern  in Fig.~\ref{BTObasic}d.  Such an unrelaxed film is referred as FeSe$^{BU}$ hereafter. Consistently, the Fermi surface topology  only contains orthogonal ellipses.
Based on the calculated Luttinger volume, there are  0.12 $e^{-}$ excessive electrons per Fe for all these films, similar to those  SLF on  NSTO as well.

Fig.~\ref{dispersion} presents the band dispersions for various SLF films grown on NBTO or NSTO.
 Fig.~\ref{dispersion}a,b gives the dispersions across $\Gamma$~(0,0). The generic  features for all films include a parabolic band, and a relative flat band at higher binding energy.
For films grown on NBTO, the larger $a$ correspond to a larger band mass of the parabolic band and smaller  separations between the two bands (Fig.~\ref{dispersion}b(i),(ii)), indicating stronger correlation effects with expanded lattice. The same holds true for the two films grown on NSTO (Fig.~\ref{dispersion}b(iii),(iv)).
 Intriguingly, even though both the NBTO/KTO and NSTO/KTO were terminated with the same TiO$_2$ layers that have the similar amount of oxygen vacancies and same lattice constant of 3.989~\AA~, there are clear differences in the band dispersions for FeSe$^{SX}$ and FeSe$^{BU}$ (Fig.~\ref{dispersion}b(i),(iii)).   The bands are flatter for SLF on NSTO  than for  SLF on NBTO.
As summarized in Fig.~\ref{dispersion}f,g, the two characteristic quantities (band mass and band separation) fall on two separate curves as a function of $a$, which clearly indicates the substrates have a nontrivial effects on the correlation and electronic structure in the FeSe film  on them.
The  substrates affect the bands around M~$(\pi,0)$ as well (Fig.~\ref{dispersion}c,d).  For the large $a$=3.989~\AA~, two electron bands are resolved for FeSe$^{BU}$ and FeSe$^{SX}$  but with different separations (Fig.~\ref{dispersion}d(i),(iii)) \cite{PengSTOKTO}, while for FeSe$^{BR}$ and FeSe$^{S}$, the two bands cannot be resolved \cite{TanFeSe}. However, the bandwidths of the electron bands are similar in all these films.

The superconducting properties of the FeSe films grown on the NBTO films are examined in Fig.~\ref{Tdep}.
Fig.~\ref{Tdep}a shows the symmetrized energy distribution curves (EDC's) at low temperatures along a cut across the Fermi surface of the unrelaxed FeSe domains of   FeSe$^{B}$.
The dispersion exhibits a characteristic bending back behavior after passing the Fermi momentum. This is a hallmark of   Bogliubov quasiparticle dispersion, and indicates that the gap is due to Cooper pair formation.
The symmetrized EDC's at the normal state Fermi momenta could be used to identify the gap by the spectral weight suppression at  $E_F$, which is minimally affected by  the temperature broadening effects \cite{Norman}.    Fig.~\ref{Tdep}b presents the temperature dependence of the   symmetrized EDC's  at the Fermi momentum for the unrelaxed domains in FeSe$^{B}$, and the gap decreases with increasing temperature and eventually closes above 73~K. Fig.~\ref{Tdep}c-d shows the gap behavior of  FeSe$^{BR}$, which closes above 68~K. The data for  FeSe$^{BU}$ shows rather weak coherence peaks, but with possible signs for $T_g$ above 77~K (see Supplementary Information).

The temperature dependence of the gap is shown in Fig.~\ref{Tdep}e, which could  be well fitted by the Bardeen-Cooper-Schrieffer (BCS) gap vs. temperature formula. This, together with the Bogliubov quasiparticle behavior of the dispersion, suggest that  it relates to the Cooper pair formation, although \textit{in-situ} transport is needed to check if $T_c=T_g$.  The fitted $T_g$ is 75$\pm$2~K, and 70$\pm$2~K for FeSe$^{B}$ and FeSe$^{BR}$, respectively. The former sets a new record    $T_g$ for Fe-HTS's.
Fig.~\ref{Tdep}f summarize the maximal gap versus $T_g$ for FeSe films and K$_x$Fe$_{2-y}$Se$_2$.
One can  find  a general linear trend between $T_g$  and the gap amplitude, reflecting their superconducting nature.
However, we note that
the $a$'s of the FeSe$^{BR}$ and FeSe$^{BU}$ domains differ by 5.5\%. When the lattice is expanded by such a large amount,  the antiferromagnetic superexchange interactions across the Fe-Se-Fe would enhance significantly   \cite{CaoFeSe}. In the context of the  antiferromagnetic-interaction/spin fluctuation  mediated superconductivity, which is arguably the current dominating picture for bulk Fe-HTS's  \cite{HuDing},  such a 5~K $T_g$ difference is surprisingly small.  In addition,  there is no simple relation between the effective mass and $T_g$, while the increased correlation strength usually should help to enhance the superconductivity in such a high doping regime.

\begin{figure}[t]
 \includegraphics[width=8.5cm]{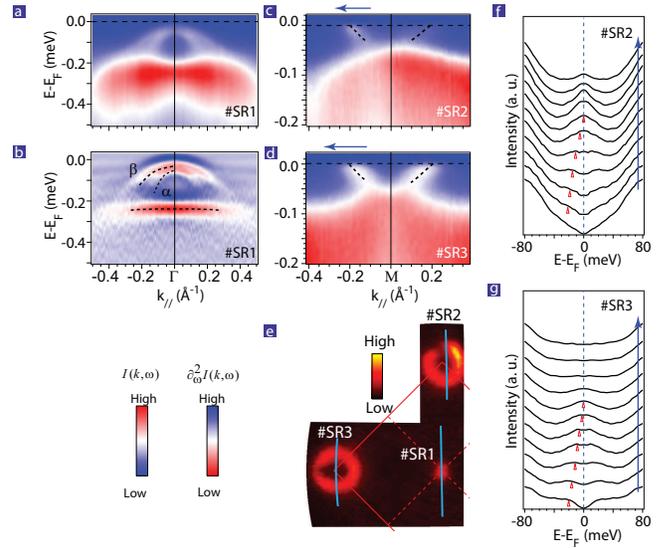}
\caption{ \textbf{Electronic structure of  FeCoSe$^{SR}$}.    FeCoSe$^{SR}$ is a Co doped thick FeSe film with relaxed lattice, $a$=3.78~\AA~.
\textbf{a,b}, The photoemission intensity (\textbf{a})  and  the corresponding second derivative with respect to energy to highlight the dispersions (\textbf{b}) along cut~\#SR1.
\textbf{c,d}, The photoemission intensity along cut~\#SR2 and cut~\#SR3, respectively.
\textbf{e}, Fermi surface map of FeCoSe$^{SR}$, where the  momentum cuts for data in other panels are marked.
\textbf{f,g}, The symmetrized EDC's  along a portion of cut~\#SR2 (\textbf{f}) and cut~\#SR3 (\textbf{g}), as indicated by the blue arrows   in \textbf{c} and  \textbf{d}, respectively. The dispersions are marked by the red triangles.   The data were taken at 30~K.}
\label{FeCoSe}
\end{figure}

The  $a$ of   FeSe$^{BR}$ is close to that of  bulk FeSe. It has been shown both by experiments and calculations that the undoped bulk FeSe has similar electronic structure as the iron pnictides with both electron and hole Fermi pockets \cite{LuFeSeCalc,TanFeSe}.
For iron pnictides,   heavy electron doping would fill up the hole pockets, bring the system to the overdoped regime, and kill the superconductivity \cite{Lifshitz1,Ye2013}.
To check if this is the case for FeSe, we have grown  35~u.c. thick cobalt-doped FeSe film on NSTO  to introduce  electron carriers (named FeCoSe$^{SR}$), and $a$ is relaxed to about 3.78~\AA~ in this film.
As shown in Fig.~\ref{FeCoSe}a,b,   the hole pockets of the $\alpha$ and $\beta$ bands sink below the Fermi energy, and the electron Fermi surface is fairly large as shown in Fig.~\ref{FeCoSe}c-e. About 8\% electrons could be doped, which is limited by the solubility of cobalt. Fig.~\ref{FeCoSe}f,g shows no signs of a superconducting gap. The $T_g$, if any, is below the lowest temperature of 30~K that could be reached at the film.
That is, for a SLF film with similar $a$ and doping, it has a high $T_g$ of $\sim$70K on NBTO, while its $T_g$ is very low if such films are staked together in FeCoSe$^{SR}$.
Had the superconductivity been originated only from the FeSe layer, one would expect the opposite: a  higer $T_g$ in FeCoSe$^{SR}$ due to  suppressed fluctuations.

Our findings thus strongly suggest that  interfacial effects could participate or even dominate the superconductivity in FeSe/NSTO or   FeSe/NBTO. For example, one could speculate some  intriguing and nonexclusive possibilities.
For example,  the  polarizability of the ions in substrates induces  interfacial electron-phonon interactions \cite{OuHongweiPRL}.  The interactions between electrons and a substrate-originated phonon  have been reported in FeSe$^S$ (ref.~\onlinecite{ZXShenFeSe}). The high phonon frequency and strong electron phonon interactions in NSTO or NBTO may mediate  superconductivity  with high $T_c$ as in BCS theory. It may also help enhance the  antiferromagnetic-interaction mediated pairing as proposed in ref.~\onlinecite{DHLee} and \onlinecite{ZXShenFeSe}.  Moreover,   Little  \cite{Little} and Ginzburg \cite{Ginzburg}  proposed almost half a century ago that the electronic polarizability of the substrate could introduce  attractive force between electrons.  This is unlikely here, since we did not find any noticeble dependence of the electronic structure and superconducting properties  on the Nb concentration (\textit{i.e.} conductivity and carrier density in the substrate).  Nevertheless, it would help to screen the  Coulomb interactions at various ranges  \cite{Sawatzky}, thus facilitate pairing \cite{Mona}. In fact, the observed substrate-dependence of  correlations in  FeSe may be resulting from the different screening of  the interactions by NSTO and NBTO.

To summarize, we have successfully fabricated FeSe films on NBTO films, and enhance the gap-closing or Cooper-pair formation temperature up to 75~K.  This establishes  interface engineering as an effective path for enhancing  $T_c$ in SLF, and pave its way towards more cost-effective  applications. More importantly,  our data suggest that   the substrate has a profound impact on the electronic structure in single layer FeSe/NSTO and   FeSe/NBTO films, and their high temperature superconductivity probably results from an extraordinary interfacial  mechanism. These results enrich the current understanding of interfacial superconductivity and high temperature superconductivity in general.

\textbf{Methods:}

\textbf{Thin film growth}

Over 40 unit cells of highly conductive 5\% Nb doped BTO films  were grown layer-by-layer on KTaO$_3$ (KTO)  substrate with ozone-assisted molecular beam epitaxy (MBE).  During the growth of Nb:BTO, the reflection high-energy electron diffraction~(RHEED) pattern retained its two dimensional character. With the shutter-controlled growth mode, the film is terminated with TiO$_2$ layer after the growth of each unit cell except for FeSe$^{BU}$ (ref. \onlinecite{RHEEDSchlom}). The  Nb:BTO films were then transferred under an ultra-high vacuum to another MBE chamber, where heat treatment were performed, and then FeSe thin films were deposited following the  procedure used in ref.~\onlinecite{TanFeSe}.
In this way,  a  FeSe/Nb:BTO/KTO heterostructure is fabricated.

\textbf{ARPES measurement}

ARPES data were taken \textit{in situ} under ultra-high vacuum of $1.5\times 10^{-11} mbar$, with a SPECS UVLS discharge lamp (21.2eV He-I$\alpha$ light) and a Scienta R4000 electron analyzer. The energy resolution is 8~meV and the angular resolution is 0.3$^{\circ}$. Data were taken at 30~K if not specified otherwise. To eliminate the photoemission charging effect due to insulating KTO, silver paste was attached on the substrate edge before growth.

\textbf{Acknowledgement:} This work is supported in part by the National Science Foundation of China under the grant Nos. 91021001 and 91221303, and National Basic Research Program of China (973 Program) under the grant No. 2012CB921402.

\textbf{Author contributions:}  R.P., H.C.X., S.Y.T., M.X., Q.S. and Z.C.Huang   grew the films, R.P., H.C.X., S.Y.T., B.P.X., X.P.S., Z.C.Huang and C.H.P.Wen performed ARPES measurements. R. P., H.C.X., S.Y.T. and D.L.F. analyzed the ARPES data.   D.L.F. and R.P. wrote the paper. D.L.F., B.P.X., and T.Z. are responsible for the infrastructure, project direction and planning.

\textbf{Additional Information:} The authors declare no competing financial interests.  Correspondence and requests for materials should be addressed to D.L.F. (dlfeng@fudan.edu.cn).


\end{document}